\input{aipcheck}

\documentclass[
    ,final            
  ]
  {aipproc}

\layoutstyle{6x9}

\def \inte {INTEGRAL}
\def \xmm {XMM--Newton}
\def \sax {BeppoSAX}
\def \src {4U~1850--087}
\def \glob {NGC~6712}

\def \hcm {\hbox {\ifmmode $ atom cm$^{-2}\else atom cm$^{-2}$\fi}}

\begin{document}

\title{INTEGRAL observations of the X--ray burster \src: 
first detection of hard X--ray emission extending above 50~keV
}

\classification{95.85.Nv 97.20.Pm 97.60.Jd 97.80.Jp}

\keywords      {X--rays; Neutron Stars; X--ray Binaries}

\author{L. Sidoli}{
address={INAF, Istituto di Astrofisica Spaziale e Fisica Cosmica, Milano, Italy}
}

\author{A. Paizis}{
address={INAF, Istituto di Astrofisica Spaziale e Fisica Cosmica, Milano, Italy}
}

\author{A. Bazzano}{
  address={INAF, Istituto di Astrofisica Spaziale e Fisica Cosmica, Roma, Italy }
}

\author{S. Mereghetti}{
  address={INAF, Istituto di Astrofisica Spaziale e Fisica Cosmica, Milano, Italy} 
}

\begin{abstract}
The X--ray burster 4U1850--087, located in the globular cluster NGC~6712,
is an ultracompact binary likely harbouring a degenerate companion.
The source has been observed with INTEGRAL several times, during the
monitoring of the Galactic plane, with an unprecedented exposure time.
The broad-band spectrum (2-100 keV; INTEGRAL together with
a quasi-simultaneous XMM-Newton observation) is well described with a disk-blackbody emission
(kT$_{in}$=0.8~keV) together with a power-law (photon index of 2).
We report here the first detection of hard X--ray emission
from this source above 50 keV. A  lower limit on the presence
of a high energy cut-off can  be placed at E$_{c}$$>$100~keV.
\end{abstract} 

\maketitle


\section{Introduction}

\src\ is an X--ray burster
located in the galactic globular cluster
\glob, associated with an optical counterpart which displays an UV
modulation \cite{Anderson1993} with a period of
20.6 minutes. If interpreted as the orbital period, this 
implies a degenerate companion of 0.04$M_\odot$ \cite{Homer1996}. 
The first broad-band spectrum 
was  obtained with 
\sax\ in the energy range 0.3--50~keV 
in 1997 \cite{SidoliGLOBS2001}, with an estimated 
0.1--100~keV luminosity of 1.9$\times10^{36}$~erg~s$^{-1}$ (assuming a distance of 6.8~kpc).
Other X--ray observations below 10~keV have been performed with 
$ASCA$ \cite{Juett2001}, \xmm \cite{SidoliLOP2005},
 $Chandra$ \cite{JuettC2005}.
We discuss here the first detection of \src\ above 50~keV with the \inte\ satellite (the details
are reported in \cite{SidoliPBM2006}).

\section{Observations and Results}

The ESA $INTEGRAL$ gamma-ray observatory was launched in October 2002 and
carries three co-aligned coded mask telescopes: the imager IBIS
\cite{Ubertini2003}, which allows high angular resolution
imaging over a large field of view (29$^{\circ}\times29^{\circ}$)
in the energy range 15\,keV--10\,MeV, the spectrometer SPI
(\cite{Vedrenne2003}; 20 keV--8\,MeV) and the X--ray monitor JEM-X
(\cite{Lund2003}; 3--35\,keV).


We analyzed 909 individual pointings (Science
Windows, SWs) within 10$^{\circ}$ from the source position, 
performed between March 2003 and November 2005 during the Galactic plane monitoring.
The version 5.1 of the OSA \inte\ 
analysis software has been used to process the data. 
\src\ is a faint source, therefore 
a meaningful spectral analysis has been performed
adding together several observations. 
Thus we grouped the pointings in four data-sets (listed
in Table~1), from which we extracted four different IBIS/ISGRI spectra,
adopting the standard spectral extraction method.

\vspace{1.5cm}
\begin{table}[ht!]
\caption{Summary of all \inte\ observations analysed here.
Four data-sets have been considered for brevity.
We list the Start and Stop Time
of the four groups of observations (3rd and 4th columns), together
with the number of SWs in each data-set (5th column).
}
\begin{tabular}[c]{lllll}
\hline
\hline\noalign{\smallskip}
Data  & Temporal  & Start Time             &  End Time   &  Num of    \\
set          &  window & (MJD)                 & (MJD)          &  SWs    \\
\noalign{\smallskip\hrule\smallskip}
       1 &  Mar 2003--May 2003  &   52708.7  &        52772.0  &   258  \\
       2 &  Sep 2003--Nov 2003  &   52910.9  &        52963.4  &   210  \\
       3 &  Mar 2004--May 2004  &   53075.1  &        53126.5  &    84  \\
       4 &  Aug 2004--Nov 2005  &   53238.1  &        53684.0  &   357 \\
\noalign{\smallskip\hrule\smallskip}
\end{tabular}
\end{table}




We fit independently the four IBIS/ISGRI spectra with a single power-law, resulting
in a photon index in the range 2-2.7 (90\% confidence level),  with
no evidence for a long-term spectral changing; the source flux (20--100 keV)
varied within a factor $\sim$2 (from 0.8$\times$10$^{-10}$~erg~cm$^{-2}$~s$^{-1}$ to
1.8$\times$10$^{-10}$~erg~cm$^{-2}$~s$^{-1}$).

We  extracted also JEM-X spectra, but the smaller field of view, 
the faintness of the source, together with 
some instrumental issues, severely reduced the
number of useful observations.
Thus data could be used only for the data-set 3 and part of the data-set 4,
with a net exposure of 76.7~ks and 55.5~ks 
respectively in the two JEM-X units (5--20~keV).
When combined together, the JEM-X plus IBIS/ISGRI 
simultaneous spectrum (data-set 3) is 
well fitted with an absorbed  power-law 
(column density fixed at 2.5$\times$10$^{21}$~cm$^{-2}$, the
expected interstellar absorption towards the globular cluster \glob) with
a photon index of 2.2$\pm{0.2}$. 
The 5--100~keV flux corrected for the absorption
is (2.4$\pm{0.3}$)$\times$10$^{-10}$~erg~cm$^{-2}$~s$^{-1}$ 
(based on the IBIS/ISGRI response matrix, $\chi^2$/dof=27.9/23).

\src\ has been
observed with \xmm\ on 27 September 2003 \cite{SidoliLOP2005}, i.e. within the period covered
by \inte\ data-set 2, with the main aim
to study the low energy absorption intrinsic to the source. 
We combined EPIC PN and IBIS spectra in order to   interestingly 
extend the spectral study of this source in the soft X--rays below 5~keV.
The best-fit resulted in a disk-blackbody emission together with a power-law 
(see Fig.~1 for the energy spectrum),
with the following parameters: an absorbing column density of
(4$\pm{2}$)$\times$10$^{21}$~cm$^{-2}$, a photon index of 2.07 $^{+0.07} _{-0.15}$,
a disk-blackbody temperature, kT$_{in}$, of 0.8$\pm{0.1}$~keV, and a inner disk
radius of r$_{\rm in}$$\times$$(cos(i))^{0.5}$=1.7$^{+1.1} _{-0.4}$~km (at 6.8~kpc).
The source flux (2--100 keV) corrected for
the absorption is 2.8$\times$10$^{-10}$~erg~cm$^{-2}$~s$^{-1}$
(based on the EPIC pn response matrix). 
Adopting a cut-off power-law instead of a power-law
in the two-component model ($\chi^2$/dof=172.2/202), 
the cut-off energy resulted in E$_{c}$$>$110~keV (90\% confidence level), 
with a best-fit photon index of 1.9$\pm{0.1}$.


\begin{figure}
  \includegraphics[height=.45\textheight,angle=-90]{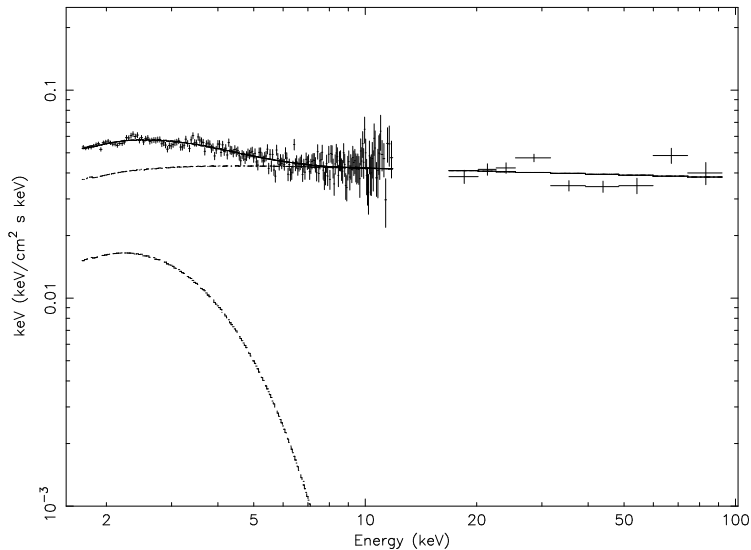}
  \caption{Broad-band \src\ EPIC/PN spectrum (obtained in September 2003) 
combined with the quasi-simultaneous IBIS/ISGRI (data-set 2),
fitted with a disk-blackbody plus a power-law.
}
\end{figure}


\section{Discussion and Conclusions}

\src\ was previously observed at high energies only with \sax\ in April 1997,
up to 50~keV \cite{SidoliGLOBS2001}.
Here we report the discovery of hard (50--100~keV) 
X--ray emission from  \src\ and a long-term study of its X--ray 
spectral behaviour.
In order to properly compare  \inte\ and  \sax\ high energy spectra,
we re-analysed the \sax\ observation 
adopting the same model used here, 
a disk-blackbody emission together with a power-law.
This resulted in 
an absorbing column density of (0.46$\pm{0.03}$)$\times$10$^{22}$~cm$^{-2}$,
an inner disk blackbody temperature, kT$_{\rm in}$, of 0.66$\pm{0.03}$~keV,
and a powerlaw photon index of 1.96$\pm{0.06}$, in very good 
agreement with the \inte\ spectroscopy.


\begin{figure}
  \includegraphics[height=.45\textheight,angle=-90]{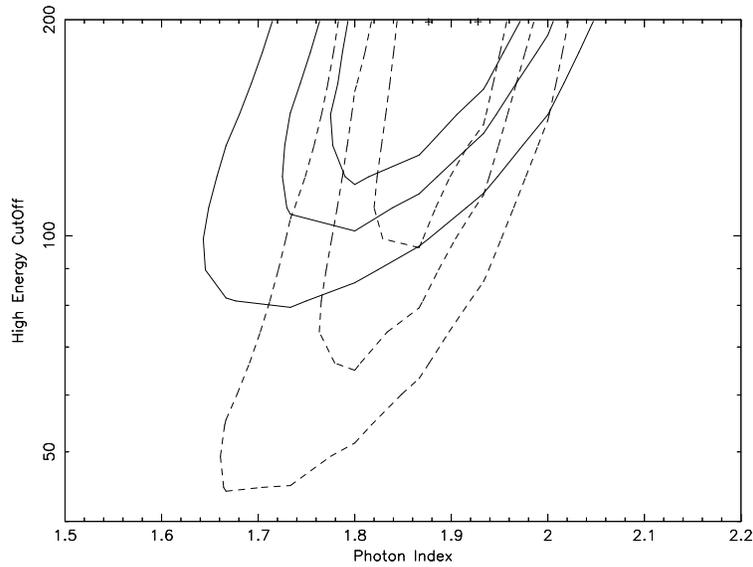}
  \caption{Comparison of the confidence contour levels for the
high energy cutoff (in units of keV): solid contours 
have been derived analysing EPIC--IBIS joint spectrum,
while the dashed contours mark the \sax\ results.
}
\end{figure}

The new observations with \inte\ interestingly allow us to put much more stringent 
limits on the presence of a high energy cut-off.
The confidence contour levels for the high energy cut-off and the power-law photon index
are compared in Fig.~2.
During the \sax\
observation a lower limit to the high energy cut-off 
could be placed at E$_{c}$$>$60~keV (90\% level),
while now with \inte\ we can interestingly shift it towards much higher energies, with  E$_{c}$$>$100~keV.
This allows us to conclude that \src\ belongs to the class of the hardest type I X--ray bursters
in our Galaxy, and the analysis of the four different \inte\ spectra suggests that the source spends
most of its lifetime in this hard spectral state (on timescales of months or years, if compared with \sax).


\begin{theacknowledgments}
We acknowledge the Italian Space Agency financial and programmatic support via contract I/R/046/04.
\end{theacknowledgments}



\bibliographystyle{aipproc}   

\bibliography{biblio}

\IfFileExists{\jobname.bbl}{}
 {\typeout{}
  \typeout{******************************************}
  \typeout{** Please run "bibtex \jobname" to optain}
  \typeout{** the bibliography and then re-run LaTeX}
  \typeout{** twice to fix the references!}
  \typeout{******************************************}
  \typeout{}
 }

\end{document}